\documentclass[preprint]{aastex}
\shorttitle{Delay/Delay-Rate Filters}
\shortauthors{Parsons and Backer}

\usepackage{amsmath}
\usepackage{graphicx}
\usepackage{natbib}
\newcounter{USEEPS}
\begin{document}
\title{Calibration of Low-Frequency, Wide-Field Radio Interferometers
Using Delay/Delay-Rate Filtering}
\author{Aaron R. Parsons and Donald C. Backer}
\affil{Astronomy Department \& Radio Astronomy Laboratory,\\
    University of California, Berkeley, CA 94720-3411}
\email{aparsons@astron.berkeley.edu}

\begin{abstract}
We present a filtering technique that can be applied to individual baselines of
wide-bandwidth, wide-field
interferometric data to geometrically select regions on the celestial sphere
that contain primary calibration sources.  The technique relies on the Fourier
transformation of wide-band frequency spectra from a given
baseline to obtain one-dimensional ``delay images'', and then the transformation of
a time-series of delay images to obtain two-dimensional
``delay/delay-rate images.'' Source selection is possible in these images
given appropriate combinations of baseline, bandwidth, integration time and
source location. Strong and persistent radio frequency interference (RFI)
limits the effectiveness of this source selection owing to the removal of
data by RFI excision algorithms.  A one-dimensional,
complex CLEAN algorithm has been developed to compensate for RFI-excision
effects. This approach allows CLEANed, source-isolated data to be used to
isolate bandpass and primary beam gain functions.
These techniques are applied to data from the
Precision Array for Probing the Epoch of Reionization (PAPER) as a
demonstration of their value in calibrating a new generation of
low-frequency radio interferometers with wide relative bandwidths and large
fields-of-view.  
\end{abstract}


\section{Introduction}
\label{sec:intro}

The rapid growth of the capabilities of digital signal processing is enabling
a new generation of interferometric arrays based on large numbers of antennas
and/or wide instantaneous frequency coverage.  Current examples of such arrays
include 
the Expanded Very Large 
Array\footnote{http://www.aoc.nrao.edu/evla} (EVLA), 
the Allen Telescope Array\footnote{http://ral.berkeley.edu/ata} (ATA), 
the LOw Frequency ARray\footnote{http://www.lofar.org} (LOFAR), 
the Precision Array for Probing the Epoch of 
Reionization\footnote{http://astro.berkeley.edu/\~{}dbacker/eor} (PAPER), 
the Murchison Widefield Array\footnote{http://haystack.mit.edu/ast/arrays/mwa} 
(MWA),  
the Long Wavelength Array\footnote{http://lwa.unm.edu} (LWA), 
the Karoo Array Telescope\footnote{http://www.kat.ac.za} (KAT), 
and the Australia Square Kilometer Array 
Prototype\footnote{http://www.atnf.csiro.au/projects/askap} (ASKAP).  
The increase in the number of
elements in these arrays is a result of requirements for larger collecting
areas and a trend towards smaller individual antenna elements.  This trend
reflects an evolving curve of array cost versus antenna element size
\citep{weinreb_daddario2001} whose minimum is shifting towards smaller antennas
(with diameter ${\rm d}$) as the cost of array correlation 
(scaling as $\sim$${\rm d}^{-4}$)
drops relative to the materials cost of producing the array antennas (scaling
approximately as $\sim$${\rm d}^{\ 0.7}$).

The smaller individual elements of new arrays have larger fields-of-view (FoVs)
that result in faster surveying speeds. However, when many parameters are
poorly known, a larger FoV complicates early array calibration by decreasing
the extent to which a single source dominates the correlated flux between
antennas.  Without isolation of sources, self-calibration cannot be performed
as a direct computation using raw data, but rather must rely on a priori models
of the sky and primary beam response pattern to divide out the
baseline-dependent interference pattern \citep{cornwell_fomalont_1989}.  This
interference pattern is illustrated by the sum over sources in the basic
measurement equation for interferometric response to a set of point sources:
\begin{equation}
V_{ij}(\nu,t) = G_{ij}(\nu, t)\sum_n{
    A_{ij}(\nu,\vec{s}_n(t))S_{n}(\nu)
    e^{2\pi\nu i(\tau_{g,ijn}(\nu,t) + \tau_{e,ij})}
}
\label{eq:meq}
\end{equation}
where $i,j$ are antenna indices, $\nu$ is radio frequency, $t$ is time, $G$ is
the complex frequency-dependent electronics gain, $A$ is the antenna beam gain
in the source direction $\vec{s}$ with unit normalization toward the zenith, 
$S$ is
the source flux, $\tau_g$ is the geometric delay for baseline $i,j$ in the
direction $\vec{s}_n$, and $\tau_e$ is the non-geometric, relative electrical 
path delay.  Both $G$,$A$ and  $\tau_g$,$\tau_e$ are typically factorized 
into antenna-based gains and delays, respectively.
Accurate solutions for the internal degrees of freedom in an array, 
especially for such parameters as the spatial variation of antenna beam gains, 
requires access to a variety of calibration sources.  Wide-field
arrays have the problem that their FoV nearly always includes sources so
bright that their sidelobes conceal lesser sources useful for
calibration.  The need to remove sources whose phase and amplitude solutions
are not of interest in order to access sources that are
useful for calibration can be a time-consuming and distracting process.

However, advances in feed design and processing bandwidth are also increasing
the amount of frequency data available in the latest interferometers.
Wide-bandwidth data with sufficient channel resolution make the delay 
transform--the Fourier transform of a frequency spectrum--a
powerful tool defining delay patterns (see \S2.2 of \citet{thompson_et_al2001})
that separate sources on the sky.  In the following sections,
we discuss techniques for using the delay (D) transform and its analog along
the time axis, the delay-rate (DR) transform, to separate the fluxes of strong
celestial sources.  Through the construction of delay/delay-rate (DDR) filters,
we demonstrate how sources may be isolated to facilitate self-calibration in
wide-FoV, wide fractional bandwidth interferometers.  Following the development
of these calibration techniques, we demonstrate their application to data from
PAPER \citep{bradley_et_al2005}, a low-frequency, non-phase-tracking, dipole
array whose steradian FoV and 100-MHz bandwidth motivated this work.

\section{The Delay (D) Transform}
\label{sec:delay}

The frequency spectrum of visibilities on a measured baseline as a function
of time reflects an interference pattern between the complex vectors
corresponding to each coherently added point source in the primary beam (Eq.
\ref{eq:meq}).  As discussed in the previous section, the calibration process
may be significantly simplified provided that the several interfering
components present in the data of each channel may be separated from one
another.  This so-called ``source separation'' should ideally be tunable in its
precision, so that coarse separation may be achieved using poorly
characterized parameters, and increasingly accurate calibration improves the
achievable separation.  For a single baseline, there are only two parameters
available for separating sources: frequency and time.  In the next two
sections, we will discuss techniques for using both of these parameters to
separate the flux of point sources.

\begin{figure}
\begin{center}
\ifnum\value{USEEPS}>0%
\includegraphics[scale=.9]{dly_fng_contours.eps}%
\else%
\includegraphics[scale=.9]{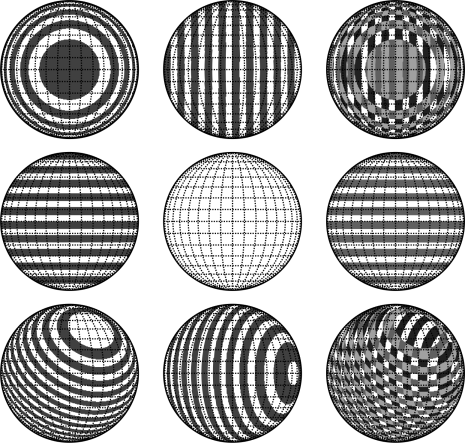}%
\fi
\caption{%
Plotted above are (left to right) delay, delay-rate, and combined contours
projected onto a celestial hemisphere centered on zero hour angle (horizontal
axis) and zero declination (vertical axis).  These contours illustrate regions
of constant width in delay/delay-rate (DDR) coordinates for baselines that are
oriented (respectively, top to bottom) in the equatorial plane pointing out 
of the page, parallel to the polar axis, and $25^\circ$ west of north
and tangent to $40^\circ$ S latitude.  DDR filters select for
intersections between these contours (third column).  The width of a selected
region depends on the orientation of a baseline with respect to a source,
and for most orientations, there exists another ``lobe'' of sensitivity
where isocontours of delay and delay-rate intersect for a second time.
\label{fig:dly_fng_cont}}
\end{center}
\end{figure}

Within the spectrum of a baseline at a given time, each source exhibits a
linearly varying phase versus frequency, reflecting the geometric group delay
associated with the projection of the baseline in the direction of
the source.  
\begin{equation}
  \tau_g(\nu,t) \equiv
    \frac{b_x}c\cos\delta\cos H(t)
    -\frac{b_y}c\cos\delta\sin H(t)
    +\frac{b_z}c\sin\delta
  \label{eq:dly}
\end{equation}
where $(b_x,b_y,b_z)$ are baseline components with units of length in 
the radial, eastern, and northern polar directions, respectively, $\delta$ is
the source declination, and $H\equiv h-\alpha$ is the source hour angle as a
function of sidereal time $h$.  The geometric delay is frequency-independent,
and can be extracted using a Fourier transform between frequency (F)-domain
and delay (D)-domain coordinates:
\begin{equation}
  \begin{aligned}
    \hat V_{ij}(\tau,t)&=\int_{-\infty}^{\infty}{
      G_{ij}(\nu,t)\left[\sum_n{
        A_{ij}(\nu,\vec s_n(t)) S_n(\nu)
        e^{2\pi\nu (\tau_{g,ijn}+\tau_{e,ij})}
      }\right]
        e^{-2\pi i\nu\tau}\ d\nu
    }\\
    &=\hat G_{ij}(\tau,t)
    \ast\sum_n{\left[
        \hat A_{ij}(\tau,\vec s_n(t))
        \ast
        \hat S_n(\tau)
        \ast
        \delta_D\left(\tau_{g,ijn} +\tau_{e,ij}-\tau\right)
    \right]}
    \label{eq:dly_t}
  \end{aligned}
\end{equation}
As illustrated above, this D transform maps the flux from each
interfering source to the corresponding delay, which will typically include a
systematic, non-geometric delay $\tau_{e,ij}$ owing to the relative
electrical signal path delays
between antennas $i$ and $j$.  While this procedure is effective as a first
step in source separation, the D transform does not result in a one-to-one
mapping of the celestial sphere to delay coordinates; sources that lie in a
plane perpendicular to the baseline vector share the same geometric delay (see
Fig. \ref{fig:dly_fng_cont}).  Furthermore, 
frequency-dependent
interferometer gains create a convolution kernel that spreads the gain of a
source in D domain, resulting in an effective delay resolution.  For a
flat passband, this resolution is approximately related to the bandwidth
sampled by the interferometer $\Delta\tau\sim\frac1{\Delta\nu}$, and translates
to a ring of finite width at the intersection of the celestial sphere with a
plane of constant delay.  Given the finite D-domain resolution, we will
hereafter assume that $\hat V_{ij}(\tau,t)$ is sampled in ``delay bins'' of
width $\Delta\tau$.

The ring on the sky defined by a delay bin centered on 
$\tau_0=\tau_{g,ij} +\tau_{e,ij}$ can be translated into coordinates of
right ascension and declination $(\alpha,\delta)$ using Eq. \ref{eq:dly}.
In celestial coordinates, the width and orientation of a 
delay ring evolve with time, as expressed using partial derivatives
of Eq. \ref{eq:dly} with respect to $\alpha$ and $\delta$:
\begin{equation}
  \begin{aligned}
  \Delta\tau=&\frac{\partial\tau_g}{\partial\alpha} \Delta\alpha+
            \frac{\partial\tau_g}{\partial\delta} \Delta\delta\\
      =& -\left[\frac{b_x}c\cos\delta\sin H
          +\frac{b_y}c\cos\delta\cos H\right]\Delta\alpha\\
        & +\left[-\frac{b_x}c\sin\delta\cos H
          +\frac{b_y}c\sin\delta\sin H
          +\frac{b_z}c\cos\delta\right]\Delta\delta
    \label{eq:dly_prm}
  \end{aligned}
\end{equation}
Given the time-variable orientation of the delay ring containing
the fixed point on the celestial sphere $(\alpha_0,\delta_0)$, 
and looking ahead to the next section where we will
show how the flux in a delay ring can be further localized using 
delay-rates, we will ignore the parametric nature of a delay ring and simply
compute an average full-width in $ \Delta\alpha, \Delta\delta
$ corresponding
to $\Delta\tau$ by averaging Eq. \ref{eq:dly_prm} over 
$H_0=h-\alpha_0\in [-\frac\pi2,\frac\pi2)$:
\begin{equation}
  \begin{aligned}
    \left<\Delta\alpha\right>=
    &\Delta\tau/\left|\frac{2b_y}{\pi c}\cos\delta\right|\\
    \left<\Delta\delta\right>=
    &\Delta\tau/\left|-\frac{2b_x}{\pi c}\sin\delta+\frac{b_z}c\cos\delta\right|
    \label{eq:dly_wid}
  \end{aligned}
\end{equation}
The angular width $\theta_D$ of a delay filter may be approximated to
order-of-magnitude as
\begin{equation}
    \theta_D ({\rm deg})\sim
    \frac{30}{\Delta\nu ({\rm MHz}) \left|\vec b\right| ({\rm km})},
\end{equation}
where $\Delta\nu$ represents the bandwidth used in the delay transform and
$\left|\vec b\right|$ is the length of the baseline involved.
For a 100-MHz bandwidth and a 1-km baseline, a delay filter has a resolution
of approximately 20 arcminutes.

%
%

\section{Delay-Rate (DR) Filtering}
\label{sec:fringe}

The delay of a source on the celestial sphere changes with time owing
to the rotation of the $b_x,b_y$ components of a baseline with the Earth.  
From Eq. \ref{eq:dly}, the 
rate-of-change of delay 
of a source at $(\alpha_0,\delta_0)$ is given by:
\begin{equation}
  \frac{\partial\tau_g}{\partial{t}}=
    -\omega_\oplus \left(\frac{b_x}c\sin H_0 
    + \frac{b_y}c\cos H_0\right)\cos\delta_0
  \label{eq:fng_rate}
\end{equation}
where $\omega_\oplus$ is the rotation rate of the Earth.  By phasing 
visibilities to a source with time variable phase $2\pi\nu\tau_g(t)$
using the current best calibration parameters, 
it is possible to stop the fringe of a source so that 
a Fourier transform of the time axis over the interval $[t_0,t_1]$ 
will add visibilities coherently into an area near zero delay-rate:  
\begin{equation}
  \begin{aligned}
  \hat V_{ij}(\nu,f)
    &=\int_{t_0}^{t_1}{
    G_{ij}(\nu,t)
    \sum_n{\left[
        A_{ij}(\nu,\vec s_n(t))
        S_n(\nu)
        e^{2\pi\nu (\tau_{g,ijn}+\tau_{e,ij} -\tau_0 -\tau)}
    \right]}
    e^{-2\pi ift}\ dt}\\
    &=\hat G_{ij}(\nu,f)
    \ast\sum_n{\left[
        \hat A_{ij}(\nu,f_n)
        \ast
        \hat{S}_n(\nu)
        \ast
        \int_{t_0}^{t_1}{\delta_D(\tau_{gr,ijn}+\tau_{e,ij}-\tau_0 -\tau)
        e^{-2\pi ift}\ dt}
    \right]}
\label{eq:fng_t}
\end{aligned}
\end{equation}
A filter near zero delay-rate along the $f$ axis (the Fourier complement 
to $t$) has a resolution determined by 
$\Delta\tau\Delta f\sim\frac{\Delta\tau}{\Delta t}$ 
(the resolution of a delay filter divided by time window used in the 
DR transform).  Note that any time-variable
gain (for example, the non-tracking primary beam of a PAPER dipole)
enters as a convolution kernel along the delay-rate axis. Such a filter
restricts flux to a ring where the celestial sphere intersects the 
plane parallel to $(\frac{b_x}c\sin H_0+\frac{b_y}c\cos H_0)$ and the Earth's 
rotational axis (see Fig. \ref{fig:dly_fng_cont}).  As in the case of the 
D transform, this ring has a 
time-variable width and orientation with respect to a fixed point on 
the celestial sphere that can be described using partial derivatives 
of Eq. \ref{eq:fng_rate}:
\begin{equation}
  \begin{aligned}
    \Delta\tau\Delta f=&\frac{\partial^2\tau_g}{\partial\alpha\partial{t}} \Delta\alpha +
         \frac{\partial^2\tau_g}{\partial\delta\partial{t}} \Delta\delta\\
    =&\omega_\oplus\left(\frac{b_x}c\cos H-\frac{b_y}c\sin H\right)\cos\delta\Delta\alpha \\
      &+ \omega_\oplus\left(\frac{b_x}c\sin H+\frac{b_y}c\cos H\right)\sin\delta\Delta\delta
    \label{eq:fng_prm}
  \end{aligned}
\end{equation}
Similarly, the average full width of this filter in celestial coordinates
around the phase center $(\alpha_0,\delta_0)$ can be described by
averaging Eq. \ref{eq:fng_prm} for $H_0=h-\alpha_0\in [-\frac\pi2,\frac\pi2)$:
\begin{equation}
\begin{aligned}
\left<\Delta\alpha\right>=
    &(\Delta\tau\Delta f)/\left|\frac{2\omega_\oplus b_x}{\pi c}\cos\delta\right|\\
\left<\Delta\delta\right>=
    &(\Delta\tau\Delta f)/\left|\frac{2\omega_\oplus b_y}{\pi c}\sin\delta\right|
\label{eq:fng_wid}
\end{aligned}
\end{equation}
An order-of-magnitude estimate of the angular width $\theta_{DR}$
of a delay-rate filter is given by
\begin{equation}
    \theta_{DR} ({\rm deg})\sim
    \frac{100}{\Delta\nu ({\rm MHz}) \left|\vec b\right| ({\rm km}) \Delta t
({\rm hr)}},
\end{equation}
where $\Delta\nu$ represents the bandwidth used in the delay 
transform, 
$\left|\vec b\right|$ is the length of the baseline involved, and $\Delta t$
is the time interval used in the delay-rate transform.
Using a 1-km baseline, a bandwidth of 100MHz, and 1 hour of data,
a delay-rate filter has a resolution of approximately 1 degree.

\section{A Combined Delay/Delay-Rate (DDR) Filter}
\label{sec:combo}

By phasing visibility data for a baseline (possibly using imperfect
calibration) to a point $(\alpha_0,\delta_0)$ and performing Fourier
transformations along both the frequency and time axes, it is possible to apply
a DDR filter near the origin in delay/delay-rate space
that selects for a restricted
area near the phase center.  The fundamental resolution of this filter is
determined by the minimum bounds placed by Equations \ref{eq:dly_wid} and
\ref{eq:fng_wid}, but wider filters may be constructed by selecting multiple
bins along both the $\tau$ and $f$ axes.  After a filter has been applied
in DDR domain (either to null or extract a region near the phase
center), one can then apply the inverse Fourier transformations to return
to the frequency-time (FT) domain.  The data may then be unphased from the specified
source, if desired, to return to the original phase center.  This filtering
process is described using matrix operator notation as:
\begin{equation}
V_{ij}^\prime(\nu,t)=\phi^{-1}(\nu,t)\cdot F_{\nu\tau}^{-1}\cdot F_{tf}^{-1}
\cdot G(\tau,f)\cdot F_{ft}\cdot F_{\tau\nu}\cdot \phi(\nu,\tau)
\cdot V_{ij}(\nu,t)
\label{eq:filter}
\end{equation}
where $\phi$ is the phasor to a point on the celestial sphere, $F$ represents
a Fourier transform, and $G$ is a gain function representing the spatial filter
desired.

The DDR filtering process requires that the width of the 
convolution kernels associated with the 
frequency-dependent electronics gain, antenna beam gain, and source flux
spectra
in Eqs. \ref{eq:dly_t} and \ref{eq:fng_t} be small
compared to the filter width needed to isolate strong sources.
While this may often be the case for smoothly varying functions characterizing
the response of analog systems and the wide-band emission
of celestial sources, the excision of faulty data, particularly data containing
radio-frequency interference (RFI), challenges this assumption by
introducing sharp features into an otherwise smoothly varying function.
The effects of nulling data in a spectrum before constructing a ``delay image''
bear many similarities to the effects of incompletely sampling an aperture
in traditional synthesis imaging.  In fact, the effects
of constructing a delay image with an incompletely sampled passband can
be compensated for by using the same deconvolution techniques used in 
synthesis imaging.

A variant of the CLEAN \citep{hogbom_1974} algorithm, adapted 
to a complex function in one dimension for a celestial 
sky dominated by a handful of point sources,
is a particularly fast and robust algorithm for deconvolving the 
effect of passband gaps \citep{roberts_et_al1987}. 
In one-dimensional, complex CLEAN, a fraction of the largest magnitude feature
(by bin) of the ``dirty image''--the Fourier transform of the spectrum
containing nulled data--is iteratively propagated to a model deconvolved image
after being divided by the gain of the ``dirty beam''.  This model is then used
to derive residuals between the predicted dirty image and the actual one, 
and these
residuals are used as the dirty image in the next iteration.  In the DDR
imaging case, the dirty beam consists of the Fourier transform of the sampling
function reflected in the data.  This deconvolution process can be illustrated
as a modification of Eq. \ref{eq:filter}:
\begin{equation}
V_{ij}^\prime(\nu,t)=\phi^{-1}(\nu,t)\cdot F_{\nu\tau}^{-1}\cdot F_{tf}^{-1}
\cdot G(\tau,f)\cdot\tilde S_\tau^{-1}(f)\cdot F_{ft}\cdot
\tilde S_t^{-1}(\tau)\cdot F_{\tau\nu}
\cdot \phi(\nu,\tau)\cdot S(\nu,t)\cdot V_{ij}(\nu,t)
\label{eq:filter_deconv}
\end{equation}
where $S$ represents a sampling function whose singularity (having multiplied
some data by zero) makes it non-invertible, but whose effects can
nonetheless be undone in approximation, represented by $\tilde S^{-1}$, via an 
iterative one-dimensional deconvolution along the delay and delay-rate axes.  
In order for complex CLEAN to converge, it is vital that the estimated gain of
the dirty beam reproduce the phase of the main lobe of the complex kernel.  For
a mostly sampled aperture, it is a reasonable assumption that the phase of the
peak response of the kernel can be taken as the phase of the overall gain for
estimating updates to the clean image and for incorporating the final residuals
of the CLEAN process.  These residuals represent what is left after the CLEAN
process has converged to a specified tolerance.

The computational complexity of the CLEAN algorithm scales between $O(N)$ and
$O(N^2)$ with the number of data samples, depending on the number of
image-domain pixels whose magnitude exceeds the specified termination
tolerance.  Since the computational complexity of the Fourier transform
operation involved in DDR imaging scales as $O(N\ log(N))$, the relative
computational expense of CLEANing is sensitive to this tolerance.  Provided
that one specifies a termination tolerance that matches the degree to which a
few strong point sources dominate data, the additional computational expense of
the CLEAN operation is negligible.  The open-source software toolkit
Astronomical Interferometry in 
PYthon\footnote{\url{http://pypi.python.org/pypi/aipy}} (AIPY) contains an
implementation of DDR imaging, with the option of using one-dimensional complex
CLEAN along both delay and delay-rate axes to remove the effects of nulled
data.  

\begin{figure}
\begin{center}
\ifnum\value{USEEPS}>0%
\includegraphics[scale=.85]{clean1d_demo.eps}%
\else%
\includegraphics[scale=.85]{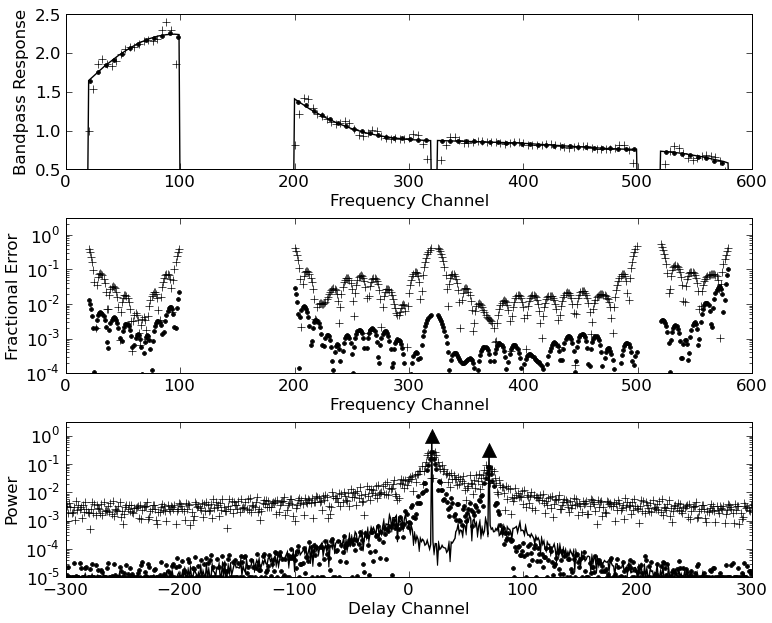}%
\fi
\caption{%
The above plots demonstrate a model application of 1-dimensional CLEAN in 
combination with delay-domain filtering to isolate a source and estimate 
passband response.  A smooth passband (solid, top) has been nulled
at various locations to reflect the removal of RFI-tainted data.  The delay
transform of this spectrum yields a ``dirty delay image'' (pluses, bottom) that
differs substantially from the model (triangles, bottom).  From this dirty
image, the maximal region around the strongest source is extracted and Fourier 
transformed back into frequency (F) domain to obtain an 
estimate of the bandpass 
shape (pluses, top and middle).  However, this estimate can be improved 
dramatically by CLEANing the dirty delay image by the delay transform 
of the sampling function (dots, bottom) before filtering to a single 
source and transforming into F domain (dots, top and middle).  Finally,
the dirty delay image may be deconvolved by the product of the sampling
function and the estimated bandpass to yield a more accurate delay image
(solid, bottom).
\label{fig:clean1d_demo}}
\end{center}
\end{figure}

A priori knowledge of the shape of the passband can be incorporated into
the model of the dirty beam to decrease the footprint of sources in the 
deconvolved image.  However, direction-dependent gains and frequency-dependent
source fluxes pose the same problems as in standard imaging 
\citep{bhatnagar_et_al2008,conway_et_al1990}--namely
that the kernels of these effects change per source, so that deconvolution
cannot be performed using a single dirty beam.
Provided that sources are sufficiently separated in DDR domain so
that the convolution kernels representing the passband, source spectrum, and
primary beam do not adversely affect source isolation, the kernels of these
functions can be used to reconstruct the corresponding FT-domain
functions.  After nulling interfering sources, deconvolving by a sampling
function, and extracting a swath in DDR domain around the source at
phase center, the remaining point-spread function reveals the effects of the
convolution 
kernels associated with the
frequency-dependent electronics gain, antenna beam gain, and source flux
spectra
in Eqs. \ref{eq:dly_t} and \ref{eq:fng_t}.  Having
preserved these functions (smoothed by a factor determined by the size of the
swath extracted), one can transform DDR data back into
FT domain to reveal their combined effect.  Given
a model source spectrum, one has direct access to each baseline's
response to that source versus time and
frequency.  This process is modeled in Figure
\ref{fig:clean1d_demo} and applied to real data in Figure \ref{fig:paper_pass}.

After using DDR filtering to isolate a source while retaining information about
the bandpass and primary beam, classic single-source self-calibration
\citep{jennison1958,pearson_readhead1984} can be used to produce the closure
phase and amplitude quantities from which antenna-based calibration parameters
can be deduced.  Because DDR filtering is sensitive to the orientation of a
baseline relative to strong celestial sources, the degree to which sources may
be isolated varies between baselines.  For this reason, it may be necessary to
manually exclude certain baselines from the self-calibration process at times
when source separation is particularly problematic.  For arrays consisting of
many more than 4 antennas, this necessity does not significantly impact the
accuracy of self-calibration.

\section{Shortcomings of DDR Filters}
\label{sec:shortcomings}

The pair of rings defined by the delay and delay-rate bins
specified in a
filter intersect at two points on the celestial sphere 
(see Fig. \ref{fig:dly_fng_cont}).  As a function of time,
one of these points of intersection remains centered on the 
specified phase center while the
other swings around the sky in a pattern that depends on the orientation of the
baseline and the location of the phase center.  
Thus, a defined filter achieves
the desired result averaged over time, but at any given time it sees two
separate areas on the sky with equal weight.  For some baseline orientations,
one of these areas can be attenuated by the primary beam, but this is not
generally the case.

The effect of having a ``double-lobed'' response has
minimal impact on source-isolation procedures. An effective
method for isolating a source consists of applying a series of narrowly
tailored nulling filters aimed at other strong sources in the field-of-view,
followed by a relatively coarse extracting filter aimed at the desired source.
This technique has the advantage of maintaining sufficient range around the
desired phase center for accommodating imperfect calibration and for 
reconstructing beam and passband shapes, while minimizing the probability
that a secondary lobe of this wide filter sweeps across another
strong source.

The complex effects of the secondary lobe of DDR filters
suggest exercising caution when constructing maps using data where such 
filters have been
used to excise certain sources.  Though such filters can indeed be used
effectively in these situations, one should remain cognizant of the fact
that an attenuating filter of changing size has been swept across wide regions
of the sky at a variable rate.  However, with increasing numbers of
baselines at unique orientations, the relative
contribution of any single secondary lobe of a DDR filter decreases.
As a result, this technique may have particularly useful applications to 
spatial imaging with large arrays of widefield antennas.

\section{Application to PAPER Calibration}
\label{sec:paper}

The Precision Array for Probing the Epoch of Reionization (PAPER) is an
experiment aimed at detecting fluctuations in 21cm emission from neutral
hydrogen at redshifts $z=7$ to $11$ as it is ionized in the first epoch of star
formation \citep{bradley_et_al2005}.  To this end, interferometric arrays of
dipole elements have been deployed at the NRAO site in Green Bank\footnote{The
National Radio Astronomy Observatory (NRAO) is owned and operated by Associated
Universities, Inc. with funding from the National Science Foundation.}, which
we call PGB, and at
the proposed Murchison Radio-astronomy Observatory (MRO) site in Western 
Australia\footnote{We acknowledge the Wajarri-Yamatji people
of Australia as the Native Title Claimants of the purposed MRO 
lands and
thank them for allowing scientific activity on the site.}, which we call PWA.
These arrays are being steadily expanded in a series of
deployments with the ultimate goal of correlating more than one hundred
antennas in the low-interference environment of Western Australia.  Currently,
PAPER has deployed a 4-element array at MRO (PWA-4) and a 16-element array in
Green Bank (PGB-16), with a typical antenna spacing of 200m.

The PAPER dipole element has a broad (125-185 MHz) frequency response owing to
a modified sleeved dipole design, with a smooth, single-lobe primary beam to
facilitate the exacting calibration that is necessary for this experiment.
Each antenna maintains a fixed pointing toward the zenith as the sky rotates
through--all PAPER data is taken as a ``drift scan''.  The full-width at
half-maximum (FWHM) of the primary beam is nearly $60^\circ$, and extends
from horizon to horizon without a null.  The wide FoV and large relative
bandwidth of the PAPER experiment have complicated progress towards an accurate
first-order calibration for the reasons outlined in \S\ref{sec:intro}.  In
particular, the RFI environment at the Green Bank site is such that wide swaths
in frequency and time must be excised before the data is usable for
astronomical purposes.  The preceding DDR filtering and CLEANing
techniques were designed precisely to combat this problem so that source
fluxes can be separated to facilitate self-calibration.

\begin{figure}
\begin{center}
\ifnum\value{USEEPS}>0%
\includegraphics[scale=.85]{paper_passband.eps}%
\else%
\includegraphics[scale=.85]{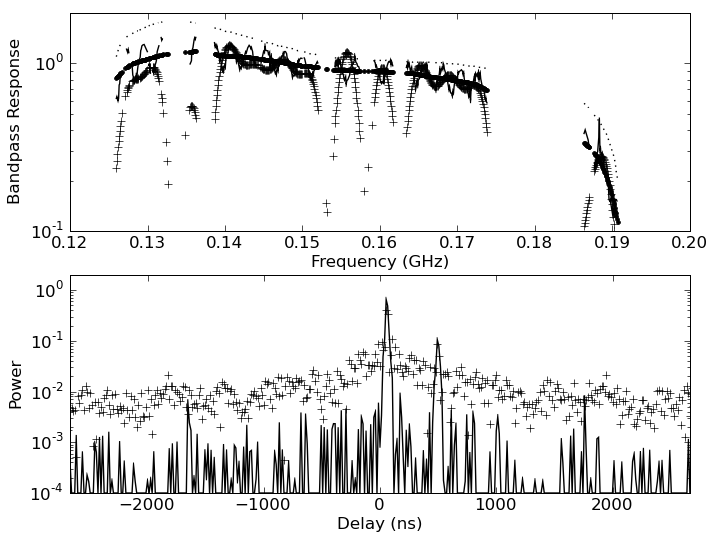}%
\fi
\caption{%
Illustrated above are the results of the same passband extraction 
process described in Figure \ref{fig:clean1d_demo}, as applied to a baseline 
spectrum recorded by the PAPER array in Green Bank, WV.  The 
magnitude of the baseline spectrum (thin solid, top) exhibits an 
interference
pattern between Cyg A and Cas A (spikes at 0, 500ns, 
respectively, in the lower plot).  Extracting a
region around Cyg A in the dirty delay image (pluses, bottom) yields a
poorly estimated passband (pluses, top).  However, CLEANing the delay
spectrum by the sampling function (solid, bottom) produces a passband estimate
(thick, top) that compares favorably with a downscaled 
auto-correlation spectrum of one of the antennas (dotted, top).  The 
difference between them is attributable to 
galactic synchrotron emission in the auto-correlation spectrum that is 
resolved out by the baseline in question.
\label{fig:paper_pass}}
\end{center}
\end{figure}

In Figure \ref{fig:paper_pass}, we demonstrate the application of a delay
filter to an integration from one baseline of PGB-16.
The substantial sidelobes in D domain that result from
RFI excision smear Cyg A and Cas A (the two dominant sources)
together, corrupting the attempt to isolate Cyg A via a delay filter
and to use that source to calibrate system gain.  By applying one-dimensional 
CLEAN
using the kernel that results from the D transform of the sampling
function, the sidelobes in D domain can be reduced dramatically.  With
cleanly separated spikes attributable to Cyg A and Cas A, 
the flux of Cyg A may be extracted with greater fidelity.  By extracting
a swath around the desired source, we preserve a convolving kernel in D
domain that retains information about
the source spectrum and smooth bandpass function.
A smooth estimate of the passband
may be obtained by Fourier transforming this kernel back into 
F domain and dividing by a known source spectrum.  Notice in Figure
\ref{fig:paper_pass} how a smooth passband resembling the auto-correlation
spectrum of one of the antennas has been constructed from the response of
Cyg A.  The difference between the auto-correlation and cross-correlation
spectrum of Cyg A is attributable to galactic synchrotron emission that is
resolved out by the baseline being used.

\begin{figure}
\begin{center}
\ifnum\value{USEEPS}>0%
\includegraphics[scale=.55]{dly_fng_map.eps}%
\else%
\includegraphics[scale=.55]{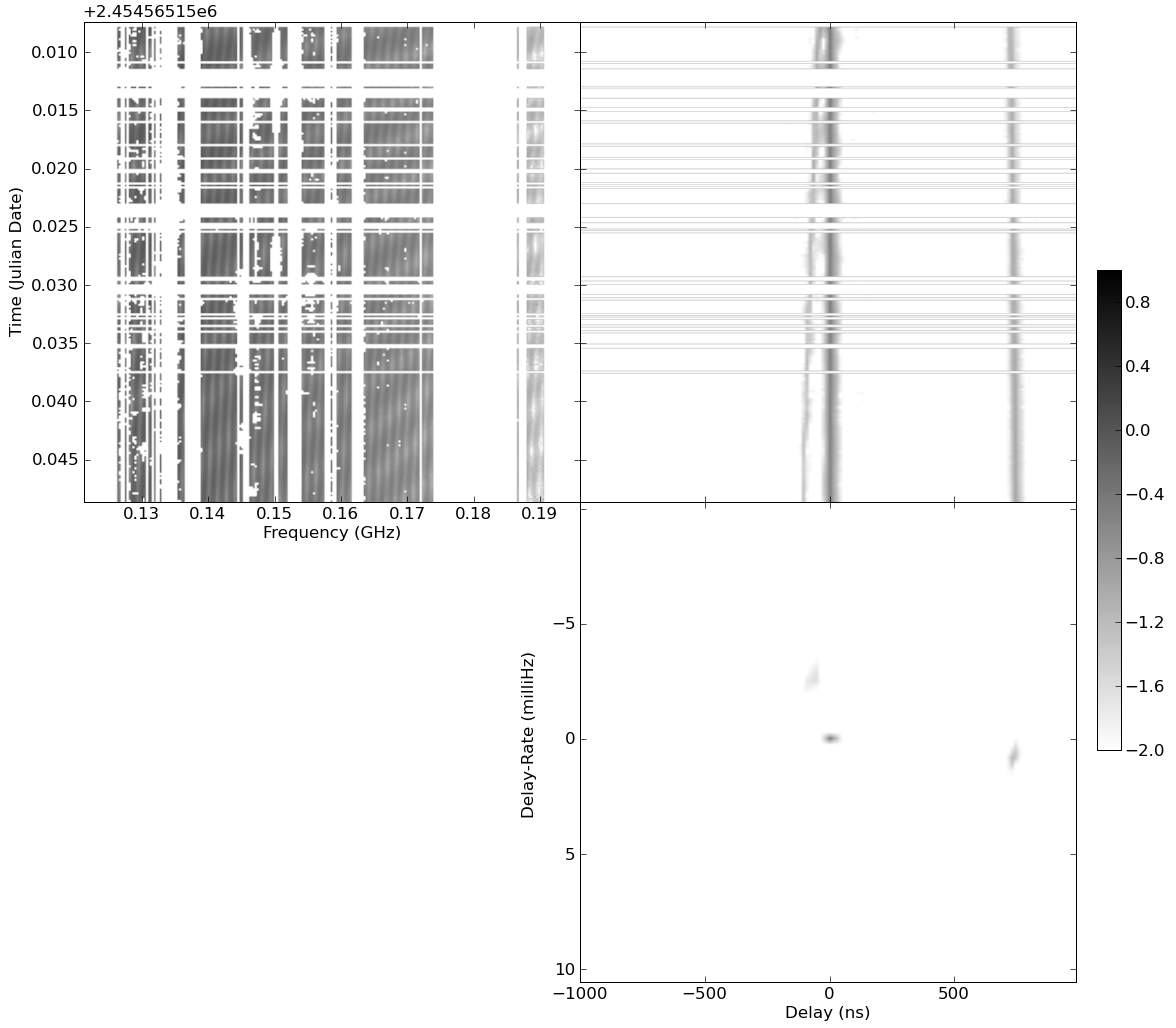}%
\fi
\caption{%
Plotted on the top-left is a 1-hour time-series of spectral data,
recorded by a single baseline of the Green Bank PAPER array, that
has been phased to
Cas A.  Beating fringes between Cas A, Cyg A, and Tau A create amplitude
ripples across the passband.
Data excised due to RFI appear as white.  The
plot on the top-right illustrates the same data, delay (D) transformed.
Flux from Cas A now forms a central stripe; Cyg A appears nearby,
drifting toward negative delay; Tau A is at high positive delay.  The effects
of flagged channels have been removed by CLEAN, but entire flagged integrations
still appear as white.  Finally, the bottom-right plot shows the result of
applying both D and delay-rate (DR) transforms.  Although nearby in D-domain,
Cas A and Cyg A now are clearly separated along the DR axis.
Cyg A and Tau A appear blurred because of their
changing DRs relative to Cas A over the period of an hour.  This
blurring illustrates why data should be phased to a region of
interest before applying DDR filters.  Note that because PAPER is a
drift-scan array, source amplitudes change with time. The effects of this are
described in \S\ref{sec:fringe}.
All of the above figures are plotted in
units of ${\rm log}_{10}$({\rm flux-density}) with a fixed 
(but arbitrary) offset.
\label{fig:dly_fng_map}}
\end{center}
\end{figure}

\begin{figure}
\begin{center}
\ifnum\value{USEEPS}>0%
\includegraphics[scale=.8]{src_sep_bm.eps}%
\else%
\includegraphics[scale=.8]{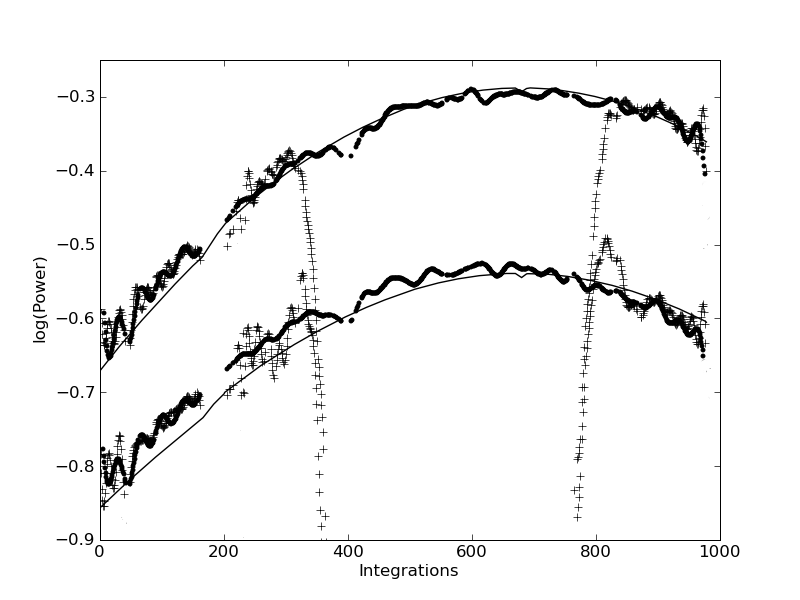}%
\fi
\caption{%
Using four hours of data from baseline described in 
Figure \ref{fig:dly_fng_map}, the above plot illustrates
the response of two channels (143 MHz on top and 165 MHz on bottom)
using data in which Cyg A as been filtered out with a narrow filter 
and Cas A has been isolated with a broader filter.  Using a strictly
delay (D)-domain filter (pluses) to remove Cyg A results in a drop-out in the
response of Cas A when the two sources cross in the D domain.  However,
a combination delay/delay-rate filter (dots) retains information about the 
smooth gain variation as Cas A drifts through the primary beam.  Predicted
models of the beam response at 143 MHz (solid top) and 165 MHz (solid bottom)
are provided for reference.
\label{fig:src_sep_bm}}
\end{center}
\end{figure}

There are cases for which pure delay filtering is inadequate for
separating sources.  The left side of Figure \ref{fig:dly_fng_map} illustrates
a waterfall plot of the delay spectrum of a baseline over the course of four
hours, during which the delay tracks of Cas A (center) and Cyg A
(right to left) cross as the sources drift through the primary beam.  In this
case, a naive D-domain filter for suppressing Cyg A and extracting
Cas A cuts a swath out of the derived spectrum for Cas A (see Fig.
\ref{fig:src_sep_bm}) and corrupts the attempted derivation of the primary beam
shape as Cas A drifts through it.  But as the right side of Figure
\ref{fig:dly_fng_map} illustrates, the DR transform provides another
axis that can be used to separate the fluxes of the sources involved.
Filtering in DDR-domain, we are able to track the
spectrum of Cas A through its intersection in the D domain with
Cyg A, and compare it to a beam model predicted through computer
simulation.  As was the case for delay filtering, we can preserve information
about changing source flux (in this case, caused by drifting through the primary
beam) by selecting a swath around the source in DR domain
containing the kernel that represents the Fourier transform of the
time-dependent gain.

\begin{figure}
\begin{center}
\ifnum\value{USEEPS}>0%
\includegraphics[scale=.7]{cas_cyg_img.eps}%
\else%
\includegraphics[scale=.7]{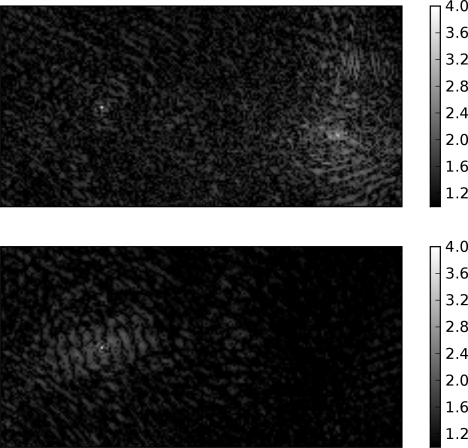}%
\fi
\caption{%
Delay/delay-rate (DDR) filters may also be used in combination with 
traditional
two-dimensional aperture synthesis imaging to null bright sources whose
sidelobes interfere with imaging other regions.  The above $45^\circ\times
90^\circ$ images of
Cas A (left, in 
${\rm log_{10}}({\rm Jy} / {\rm beam})$) illustrate how 
Cyg A (right) can be filtered out
of the data of all baselines used for imaging, resulting in the complete
suppression of the source, and an attenuation of a surrounding region of sky 
whose width is baseline-dependent.  However, the secondary lobe of
a DDR filter (\S\ref{sec:shortcomings}) causes distortions
in other regions of the image that must be weighed against the advantages
of suppressing bright sources.  In the lower panel, a grating sidelobe of 
Cas A has been introduced as the secondary lobe of one of the 
28 baselines used in imaging swept across the region.  With increasing
numbers of antennas, the relative contributions of the secondary lobes of
DDR filters decreases in imaging, making this an attractive imaging technique
for large arrays.
\label{fig:cas_cyg_img}}
\end{center}
\end{figure}

Finally, having used source isolation to self-calibrate PGB-8 data, 
we demonstrate an application of the same DDR filtering techniques to 2-dimensional 
aperture synthesis imaging.  Figure \ref{fig:cas_cyg_img} demonstrates how
a wide-field image of Cyg A and Cas A (top) can be filtered to remove
Cyg A by applying the appropriate DDR filter to data from
each baseline.  Though this technique can be very effective in removing broad
sidelobes associated with a strong celestial source, one must 
beware of the effects of the secondary lobes of these filters 
(\S\ref{sec:shortcomings}), which in this case have increased the sidelobes
associated with Cas A (Fig. \ref{fig:cas_cyg_img}, bottom).  The
secondary lobe of one of the baselines has swept across the region of
sky containing Cas A, leaving a grating interference pattern that
represents the contribution that visibilities measured by that baseline
should have made to the image, as modeled by the dirty beam used for
deconvolving the dirty 2-dimensional image.  The absence of this contribution
appears as the missing fringe pattern with an inverted sign.  Though such
filtering can prove valuable in removing strong sources to reveal weaker ones
lost in sidelobes, it is clear that the resultant images are corrupted as a
result, and that accurate imaging will require the visibilities predicted from
these corrupted images to be compared with the measured ones, possibly using
iterative sky/visibility modeling to converge on an accurate image.

\section{Conclusion}
\label{sec:conc}

We have described a technique for isolating source fluxes by Fourier 
transforming FT-ordered data from an individual baseline
into DDR domain, applying filters, and then performing the
inverse transforms.  The ability to isolate sources vastly simplifies the
calibration of interferometric arrays with wide fields-of-view where 
self-calibration would otherwise require an accurate a priori sky model
to account for multiple strong sources within the primary beam. 
Imprecise calibration and wide filters can be used initially to select wide
areas around sources that encompass calibration errors and allow improved
calibrations to be obtained.  As calibration improves, these filters can be
more narrowly tailored to allow the extraction of bandpass and beam functions
from their corresponding DDR-domain kernels, and ultimately
such filters can even be used in combination with traditional aperture
synthesis imaging to eliminate the sidelobes of strong sources and reveal
weaker celestial sources.

While such filters have many desirable qualities such as minimal reliance on
prior calibration and geometric widths that are independent of frequency,
the fact that such filters have a time-variable orientation with respect to
the celestial sphere and a secondary response lobe that sweeps across a large
area on the sky limit their usefulness for precision imaging.  However, 
combined with a modeling process that feeds estimates of the sky from
such images into model visibilities that are compared with measured data,
DDR filters may constitute an important imaging tool even for
high dynamic-range applications.  


\bibliographystyle{apj}
\bibliography{biblio}


\end{document}